\documentclass[apj]{emulateapj}
\usepackage{amsmath,amsthm,amsfonts,amssymb,bm}
\usepackage{listings}
\usepackage{color}
\usepackage{textcomp}
\usepackage{epstopdf}
\usepackage{aas_macros}
\usepackage{natbib}
\usepackage{hyperref}

\begin{document}
\title{Cosmic Shear Measurement using Autoconvolved Images}
\author{Xiangchong Li, Jun Zhang}
\affil{Center for Astronomy and Astrophysics, Department of Physics and Astronomy, \\
Shanghai Jiao Tong University, 955 Jianchuan road, Shanghai, 200240, China}
\email{betajzhang@sjtu.edu.cn}


\begin{abstract}

We study the possibility of using quadrupole moments of auto-convolved galaxy images to measure cosmic shear. The autoconvolution of an image corresponds to the inverse Fourier transformation of its power spectrum. The new method has the following advantages: the smearing effect due to the Point Spread Function (PSF) can be corrected by subtracting the quadrupole moments of the auto-convolved PSF; the centroid of the auto-convolved image is trivially identified; the systematic error due to noise can be directly removed in Fourier space; the PSF image can also contain noise, the effect of which can be similarly removed. With a large ensemble of simulated galaxy images, we show that the new method can reach a sub-percent level accuracy in general conditions, albeit with increasingly large stamp size for galaxies of less compact profiles.
 
\end{abstract}

\keywords{cosmology, large scale structure, gravitational lensing - methods, data analysis - techniques, image processing}

\section{Introduction}
\label{S-Intro}

Weak gravitational lensing refers to the small but coherent distortion of intrinsic galaxy images by inhomogeneous mass distribution \citep{bs01,hj08,kilbinger15}. Since weak lensing is a pure gravitational effect, it has a wide application in cosmology, including: probing the large scale matter distribution; constraining cosmological parameters \citep{schrabback10,kilbinger2013,fu2014,kitching2015}; determining the connection between visible galaxies and dark matter through galaxy-galaxy lensing measurement \citep{schulz10,mandelbaum16,clampitt2016}; constraining theories of gravity on cosmological scales \citep{reyes10,simpson2013,blake2016}. Precise measurement of weak lensing is one of the primary goals of several cosmological surveys (e.g. , DES\footnote{http://www.darkenergysurvey.org/} , HSC\footnote{http://www.naoj.org/Projects/HSC/}, KIDs\footnote{http://www.astro-wise.org/projects/KIDS/}, LSST\footnote{http://www.lsst.org/lsst}, WFIRST\footnote{http://wfirst.gsfc.nasa.gov/} ).

Weak lensing induces percent-level coherent shape distortions to the galaxies. It is well known that smearing due to the Point Spread Function (PSF hereafter) caused by the atmosphere and telescope is significantly larger than the weak lensing signal. In addition, the photon noise and the pixelation effect are also sources of contamination to the cosmic shear signal. It is therefore a challenging task to construct unbiased shear estimators \citep{mandelbaum14}.

Typically, quadrupole moments of observed galaxy images are used to construct shear estimator. It can be shown that a straightforward way of removing the PSF effect in shear measurement is to subtract the quadrupole moments of the PSF from those of the galaxy \citep{valdes83}. We notice that this method is not applied in practice. The most important reason is that the photon noise adds a significant contribution to the measurement of the quadrupole moments of an image, especially at large distances from the center of the source. Most commonly, an additional weight function (e.g., Gaussian) of a reasonable size is applied to the image to limit the noise contribution to the shape measurement\citep{kaiser95,rhodes01,bj02,hs03,melchior11,okura11}. Nevertheless, the existence of the weight function leads to a series of other corrections to the shear measurement that are typically hard to calculate. Furthermore, precise measurement of the quadrupole moments always requires an accurate determination of the centroid position of the source, which is difficult in the presence of noise and finite pixel size. The purpose of this paper is to understand if the noise and centroid problems can be solved by extending the framework of \citet*{valdes83} along another line of thought. 

\citet*{jz08} has proposed to use quadrupole moments of the galaxy power spectrum in Fourier space to do shear estimation. It is found that the centroid problem can be avoided in the Fourier domain, and the shear bias from the background and the Poisson noise can be systematically removed \citep{zhang15} (ZLF15 hereafter). Taking the advantage of ZLF15, we consider applying the idea of \citet{valdes83} to the auto-convolved galaxy image, which is simply the inverse Fourier transformation of the galaxy power spectrum. It turns out that the PSF effect can be similarly removed by subtracting the quadrupole moments of the auto-convolved PSF from those of the auto-convolved galaxy, instead of transforming the original PSF to a Gaussian form as done in ZLF15. Both the background noise and the Poisson noise can be removed in Fourier space statistically. Moreover, this novel method has several other advantages: 1) It only needs the quadrupole moments of auto-convolved PSF instead of its full morphology; 2) The PSF image/moments can contain noise, the effect of which can also be systematically removed; 3) It involves simple and fast image processing procedures. 

In \S \ref{S-method}, we review the relation between the cosmic shear and galaxy quadrupole moments, and introduce the idea of autoconvolution in shear estimation. In \S \ref{S-test}, we test the performance of the new method in general conditions. We also study the requirement on the galaxy stamp/aperture size for avoiding shear recovery error due to the spatial extension of the galaxy profile. This is neccessary also for avoiding possible interferences from neighbouring objects in real observations. A brief summary is given in \S \ref{S-Summary}.

\section{The Method}
\label{S-method}
\subsection{Estimator with Quadrupole Moments}
\label{G_Q_M}
Setting the origin of the coordinates at the centroid of galaxy image, one can describe the weak lensing effect as a linear mapping between the intrinsic galaxy image $f_s$ and the lensed image $f_l$ as
\begin{eqnarray}\label{shear equation}
\vec{x_l} = \textbf{A} \vec{x_s} && f_l(\vec{x_l})=f_s(\vec{x_s}).
\end{eqnarray}
where ($g_1,g_2,\kappa \ll 1$)
\begin{equation}\label{shear A}
\textbf{A}=(1+\kappa)\begin{pmatrix}
1+g_1 & g_2 \\ g_2 & 1-g_1 
\end{pmatrix}
\end{equation}
The two components of the reduced shear ($g_1,g_2$) describe the stretching of galaxy images, and the convergence $\kappa$ describes a change in size and brightness for the galaxy image. For convenience in the following discussion, we set $\kappa=0$, which does not affect our conclusions. 
Assuming the intrinsic galaxy image is statistically isotropic, one can construct the shear estimator with the quadrupole moments of lensed galaxy image as \citep{weinberg08}
\begin{equation}\label{original estimator}
\begin{split}
g_1&=\frac{<Q_l^{20}-Q_l^{02}>}{2<Q_l^{20}+Q_l^{02}>}\\
g_2&=\frac{<Q_l^{11}>}{<Q_l^{20}+Q_l^{02}>}.
\end{split}
\end{equation}
in which the quadrupole moments of the lensed galaxy image are defined as ($i,j=0,1,2$)
\begin{equation}\label{qm of lensed galaxy}
Q_l^{ij}=\int{x^iy^jf_l(\vec{x})d^2x}
\end{equation}
Note that the moments can be normalized by the flux of the source galaxy.
In the presence of PSF, the shear estimator must be constructed as a function of the observed galaxy ($f_o$), which is the convolution of the lensed galaxy ($f_l$) and the PSF ($w$)
\begin{equation}\label{PSF convolution}
f_o(\vec{x_o})=\int{w(\vec{x_o}-\vec{x_l})f_l(\vec{x_l})d^2 x_l}.
\end{equation}
In this case, shear estimators can be built on the moments of the galaxy and PSF images as \citep{valdes83}  
\begin{equation}\label{Estimator Revised}
\begin{split}
g_1&=\frac{<(Q_o^{20}-Q_o^{02})R^{00}-(R^{20}-R^{02})Q_o^{00}>}{2<(Q_o^{20}+Q_o^{02})R^{00}-(R^{20}+R^{02})Q_o^{00}>}\\
g_2&=\frac{<Q_o^{11}R^{00}-R^{11}Q_o^{00}>}{<(Q_o^{20}+Q_o^{02})R^{00}-(R^{20}+R^{02})Q_o^{00}>}.
\end{split}
\end{equation}
where the moments of galaxy and PSF are defined as ($i,j=0,1,2$)
\begin{equation}\label{qm Q&R}
\begin{split}
Q_o^{ij}&=\int{x^iy^jf_o(\vec{x})d^2x}\\
R^{ij}&=\int{x^iy^jw(\vec{x})d^2x}.
\end{split}
\end{equation}
The math details are shown in Appendix \ref{App-PR}.

\subsection{Estimator with Autoconvolution}
\label{A_C_S}
\subsubsection{Property of Autoconvolution}
Real images are recorded on CCD pixels with noise present. For shear estimator of eq.(\ref{Estimator Revised}) to be useful in practice, it requires an accurate determination of the centroid of the source image, and a procedure to remove the shear bias due to noise. As pointed out by ZLF15, the image of the galaxy power spectrum can be equally used for shear measurement. More importantly, the centroid of the power spectrum is fixed in the Fourier domain, and the noise power can be removed from the source power statistically due to their distinct features in Fourier space. We are therefore strongly motivated to apply the method of eq.(\ref{Estimator Revised}) on images related to the power spectrum of the source. It turns out that the auto-convolved source images, i.e., the inverse Fourier transformation of the source power spectrum, can be used to define the moments in eq.(\ref{Estimator Revised}) for the purpose of shear recovery. \footnote{Note that historically, \citet{waerbeke97} proposed to use autocorrelation of galaxy surface brightness field for shear measurement. The details of their method are however significantly different from ours in many aspects, including PSF correction, noise treatment, and the form of shear estimator.}

The autoconvolution of an image in our case can be defined as
\begin{equation}\label{autoconvolution}
H(\vec{x})=\int h(\vec{x}+\vec{x}')h(\vec{x}') dx'^2
\end{equation}
where $h$ could represent the intrinsic galaxy $f_s$, the lensed galaxy $f_l$, the observed galaxy $f_o$, or the PSF $w$, and $H$ could represents the corresponding auto-convolved image $F_s$, $F_l$, $F_o$ or $W$.
\begin{figure}[t]
\centering
\includegraphics[width=0.45 \textwidth]{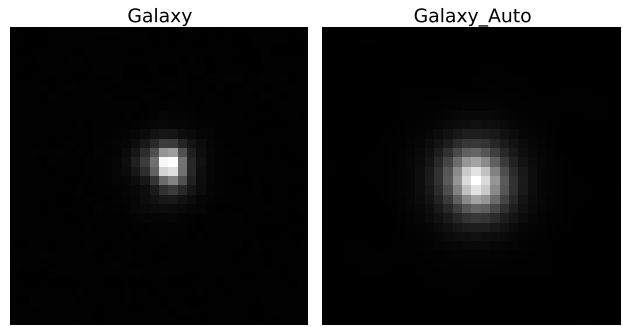}
\caption{Galaxy's surface brightness field and its autoconvolution.}\label{auto-convolved image} 
\end{figure}
The auto-convolved image has 4 good properties:\\
1) The auto-convolved image changes likes eq.(\ref{shear equation}) under the distortion by lensing
\begin{eqnarray}\label{shear auto-convolved}
F_l(\vec{x}_l)=|A|F_s(\vec{x}_s)&&
\vec{x}_l=A \vec{x}_s;
\end{eqnarray}\\
2) The autoconvolution of the observed galaxy ($F_o$) is the convolution of the auto-convolved lensed galaxy ($F_l$) and the auto-convolved PSF ($W$)
\begin{equation}\label{PSF convolution auto}
F_o(\vec{x_o})=\int{W(\vec{x_o}-\vec{x_l})F_l(\vec{x_l})d^2 x_l}.
\end{equation}\\
3) It is always symmetric to the origin (see fig.[\ref{auto-convolved image}])
\begin{equation}
H(-\vec{x})=H(\vec{x}),
\end{equation}
so the centroid of auto-convolved image is always well defined.\\
4) The auto-convolved image $H$ is the inverse Fourier transformation of the power spectrum $\tilde{H}$
\begin{equation}\label{inverse fourier transformation}
H(\vec{x})=\frac{1}{(2\pi)^2}\int \tilde{H}(\vec{k})e^{i\vec{k}\cdot\vec{x}}d^2k
\end{equation}
in which the power spectrum is defined as
\begin{equation} \label{spectral density}
\begin{split}
\tilde{h}(\vec{k})&=\int h(\vec{x})e^{-i\vec{k}\cdot\vec{x}}d^2x\\
\tilde{H}(\vec{k})&=|\tilde{h}(\vec{k})|^2
\end{split}
\end{equation}  
Therefore autoconvolution can be carried out using fast Fourier transformation.

\begin{figure}[!t]
\centering
\includegraphics[width=8cm,height=6cm]{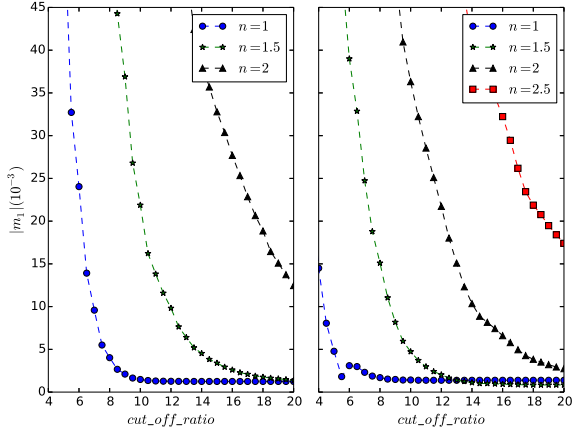}
\caption{The multiplicative biases of our autoconvolution method for galaxies of different Sersic indices. The x-axis is the ratio between the radius of the circular aperture and the half-light radius of the galaxy. The left panel shows the results for galaxies whose half-light radii are twice that of the PSF, and the right panel for galaxies whose half-light radii are similar to that of the PSF. }\label{fig_cut_A} 
\end{figure}

It turns out that the above properties allow us to construct shear estimators similar to eq.(\ref{Estimator Revised}) using the moments of $F_o$ and $W$
\begin{equation}\label{qm S&T}
\begin{split}
Q_a^{ij}=&\int{x^iy^jF_o(\vec{x})d^2x}.\\
R_a^{ij}=&\int{x^iy^jW(\vec{x})d^2x}.
\end{split}
\end{equation}
The corresponding shear estimators are
\begin{equation}\label{Estimator Centroid}
\begin{split}
g_1&=\frac{<(Q_a^{20}-Q_a^{02})R_a^{00}-(R_a^{20}-R_a^{02})Q_a^{00}>}{2<(Q_a^{20}+Q_a^{02})R_a^{00}-(R_a^{20}+R_a^{02})Q_a^{00}>}\\
g_2&=\frac{<Q_a^{11}R_a^{00}-R_a^{11}Q_a^{00}>}{<(Q_a^{20}+Q_a^{02})R_a^{00}-(R_a^{20}+R_a^{02})Q_a^{00}>}.
\end{split}
\end{equation}
The math details are shown in Appendix \ref{App-AC}.

\begin{figure}[!t]
\centering
\includegraphics[width=8cm,height=6cm]{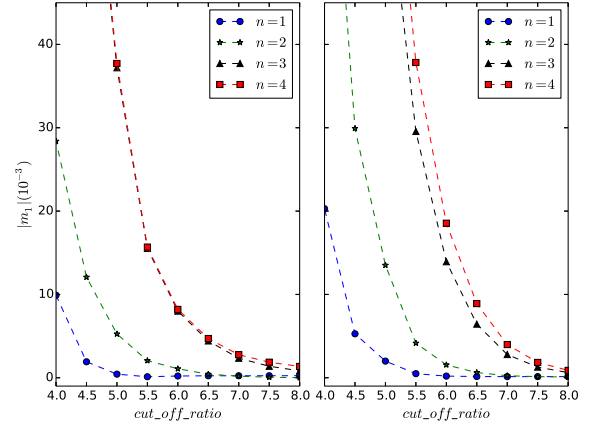}
\caption{ Same as fig.\ref{fig_cut_A}, but for the method of ZLF15. }\label{fig_cut_F} 
\end{figure}

\subsubsection{Noise Correction}

The terms on the right side of eq.(\ref{Estimator Centroid}) are contaminated by noise, including both background noise and source Poisson noise. As shown in ZLF15, the background noise contribution can be estimated and directly subtracted using a background noise image near the galaxy location, because they are not correlated with the source shapes. The source Poisson noise statistically exhibits a scale-independent power spectrum in Fourier space, therefore its contribution can be estimated at large wave-numbers, and subtracted from the source power spectrum on all scales.

These procedures can similarly be applied in our new method not only for the galaxy, but also for the PSF, owing to the fact that the new shear estimators linearly depend on the power spectra of both the galaxy and the PSF. In practice, the PSF power at the position of the galaxy can be constructed as a weighted sum of the power spectra of its neighboring stars, each of which is subtracted by the power of its companion background noise image in the neighborhood. It is in this sense that the new method allows the presence of noise in the PSF, an important feature that is not shared by ZLF15.

Let us denote the auto-convolved images of background noise for galaxy and PSF as $G(\vec{x})$ and $V(\vec{x})$ respectively. To remove the noise bias, the moments of galaxy and PSF can be redefined as
\begin{equation}
\begin{split}\label{qm Discrete}
S^{ij}=&\int{x^iy^j[F_o(\vec{x})-G(\vec{x})]d^2x}.\\
T^{ij}=&\int{x^iy^j[W(\vec{x})-V(\vec{x})]d^2x}.
\end{split}
\end{equation}
The shear estimators are updated accordingly
\begin{equation}\label{Estimator final}
\begin{split}
g_1&=\frac{<(S^{20}-S^{02})T^{00}-(T^{20}-T^{02})S^{00}>}{2<(S^{20}+S^{02})T^{00}-(T^{20}+T^{02})S^{00}>}\\
g_2&=\frac{<S^{11}T^{00}-T^{11}S^{00}>}{<(S^{20}+S^{02})T^{00}-(T^{20}+T^{02})S^{00}>}.
\end{split}
\end{equation}

\subsection{General Setup}
\label{S-test general}

The pipeline of the new method is summarized as follows\\
1) Fourier transform the galaxy and PSF images and calculate their power spectrum according to eq.(\ref{spectral density});\\
2) Remove the source Poisson noise according to ZLF15;\\
3) Inversely transform power spectrum of galaxy and PSF to get the autoconvolution according to eq.(\ref{inverse fourier transformation});\\
4) Repeat step 1) and 3) on the neighbouring background images;\\
5) Construct the shear estimator according to eq.(\ref{Estimator final})

\section{Numerical Test}
\label{S-test}

In this section, we use simulated galaxy and star images to test the accuracy of the new method under general conditions. We choose to use circular top-hat aperture to define the galaxy area for shear measurement. An important issue to address is about the convergence of the quadrupole moments as a function of radius, which determines the aperture sizes for the sources. We consider realistic galaxy and PSF profiles, and compare the results with those using the method of ZLF15 in \S \ref{S-test-cut-off}. We also test the performance of our new method and compare with ZLF15 in the presence of background and source Poisson noise in \S \ref{noise_c}.

In the following tests, each galaxy is placed on a $2^n \times 2^n$ ($n$ is integer) postage stamp. The pixel size is defined as the length unit. We use a number of points to simulate every galaxy. The advantage of point source is that one can apply image distortions (lensing, rotation, ellipticity) by directly moving the points, and can add the PSF effect by directly turning each point into a 2D profile. All these procedures can be done without doing any interpolations on the pixel-by-pixel basis \citep{jz08}. We generate point sources that are homogeneously distributed within a round disk to simulate an intrinsic face-on galaxy image. The luminosity of each points is determined by its position to form a Sersic profile \citep{sersic63} on average
\begin{equation}\label{Sersic galaxy}
f(\vec{x})=\frac{1}{a(n)r_e^2}e^{-b(n)(|\vec{x}|/r_e)^\frac{1}{n}}
\end{equation} 
where $r_e$ is the half-light radius of galaxy, $n$ is the Sersic index that determines the morphology of the galaxy; $a(n)$ and $b(n)$ are known numerical functions \citep{cb00}.
The face-on disk galaxy image is then projected onto the source plane with an inclination angle. The shape noise can be removed by using $4$ galaxies with same morphology but the intrinsic orientations are separated by $45^\circ$.
We use the Moffat profile for the PSF\citep{moffat69}
\begin{equation}\label{Moffat PSF}
w_{m}(\vec{x})=[1.+c(|\vec{x}|/r_p)^2]^{-\beta}   
\end{equation}
where $r_p$ is the half-light radius, $\beta$ typically
ranges from 2 to 5 and $c=2^{\frac{1}{\beta-1}}-1$ is a parameter.  
Generally, we quantify the shear recovery accuracy with the multiplicative bias ($m_{1,2}$) and additive bias ($c_{1,2}$) using $7$ input shears which are random numbers evenly distributed within $[-0.02,0.02]$. The multiplicative and additive bias are defined as
\begin{equation}
\label{shear_error}
g_{1,2}^{measured}=(1+m_{1,2})g_{1,2}^{input}+c_{1,2}
\end{equation}
where the subscripts $1$, $2$ refer to the first and second components of the cosmic shear respectively.
\begin{figure}[!t]
\centering
\includegraphics[width=6cm,height=6cm]{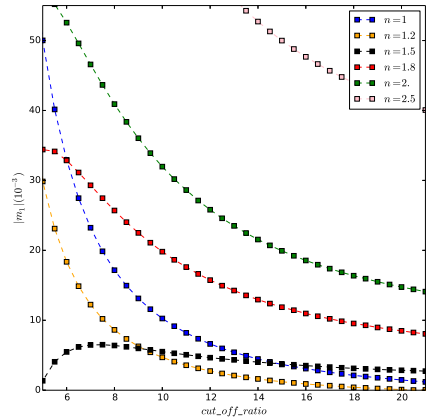}
\caption{The multiplicative biases of autoconvolution method using Gaussian filter for galaxies with different Sersic indices. The horizontal axis refers to the cut-off ratio for Gaussian filter, and the vertical axis are the multiplicative bias $m_{1}$ (upper panel) defined in eq.(\ref{shear_error}). }\label{fig_gfilter} 
\end{figure}
\subsection{Aperture radius}
\label{S-test-cut-off}
Real galaxies can have somewhat extended profiles, therefore it is necessary to study the requirement on the galaxy stamp size for avoiding shear recovery error and possible interferences from neighbouring objects in real observations. We choose to use circular top-hat aperture to define the galaxy area for shear measurement.
We define the cut-off ratio as the ratio between the radius of circular aperture and the half-light radius of the observed galaxy. We test the accuracy of our new method and ZLF15\footnote{The standard derivative of target Gaussian PSF is set to $1.5$ times of the PSF's half-light radius} under different cut-off ratio. Every galaxy and star is contained in a postage stamp of $128\times 128$ pixels. We use $4000$ points to simulate one galaxy. In our test we use $25$ galaxies with random ellipticity and each of them is rotated by $45$ degree for 3 times to remove the shape noise. The PSF has a Moffat profile with $\beta=3.5$. The PSF is truncated when radius is larger than $8$ times of its half-light radius. We compare the performance of new method and ZLF15 on two different choices of the ratio between the half-light radius of the observed galaxy image and that of the PSF. 

We plot the multiplicative bias of our new method and ZLF15 in fig.[\ref{fig_cut_A}] and fig.[\ref{fig_cut_F}] respectively. Fig.[\ref{fig_cut_A}] shows that our new method performs well for compact Sersic galaxies ($n=1.$) and the cut-off ratio should be larger than $7$ to ensure that multiplicity error is within $10^{-2}$. However, the proper cut-off ratio is strongly dependent on the galaxy's morphology. For  more extended Sersic galaxies ($n=2$), we require the cut-off ratio to be larger than $20$. As a comparison, fig.[\ref{fig_cut_F}] indicates that the performance of ZLF15 is much less dependent on the convergence rate of galaxy profile. The required cut-off ratio is only $7$ even for the most extended Sersic galaxy ($n=4$) in our examples.

For accurate shear recovery, the new method generally requires a somewhat larger stamp/filter size than ZLF15. It is interesting to ask whether Gaussian filter could improve the convergence of the autoconvolution method. We try to use the isotropic Gaussian filter to weight the galaxy image before doing autoconvolution. The cut-off ratio of Gaussian function is defined as the ratio between the RMS width of the Gaussian function and the half light radius of the observed galaxy. The galaxies and PSF's used in this experiment are the same as those of the right panel of fig.[\ref{fig_cut_A}]. Comparing fig.[\ref{fig_cut_A}] with fig.[\ref{fig_gfilter}], we find that the performance of Gaussian filter is worse than the top-hat filter. We therefore stick to the circular top-hat filter in our pipeline.

\label{S-test-NR}
\begin{figure}[!t]
\centering
\includegraphics[width=6cm,height=8cm]{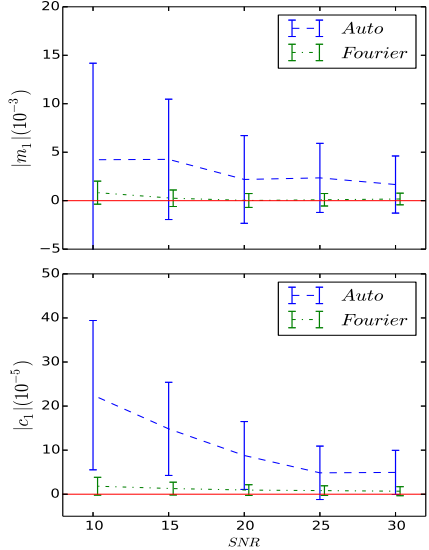}
\caption{The horizontal axis refers to the glaxy SNR, and the vertical axis are the multiplicative bias $m_{1}$ (upper panel) and the additive bias $c_{1}$ (lower panel) defined in eq.(\ref{shear_error}). 'Auto' refers to our new method and 'Fourier' refers to ZLF15. }\label{fig_noiserevision_outcome} 
\end{figure}

\subsection{Noise Correction}
\label{noise_c}

The purpose of the tests below are to test the performance of our new method and ZLF15 in the presence of noise. The postage stamps have $32\times 32$ pixels. We generate $8 \times 10^7$ different galaxies. The point sources of every galaxy are randomly placed within a disk, with luminosities assigned according to a Sersic profile with $n=0.5$ (defined in eq.(\ref{Sersic galaxy})). They are then projected onto the plane of the sky with a random angle, lensed, and convolved with the PSF. Because the number of points used in the simulation of every galaxy is small (40 points for each galaxy), the images have random ellipticities and orientations. The half-light radius of intrinsic galaxy is set to $1.4~(pix)$. The PSFs we use are Moffat with $m=3.5$ and $r_p=1.4~(pix)$, the truncate radius is $8$ times of $r_p$.

We add uncorrelated background Gaussian noise and source Gaussian noise with different Signal-to-Noise Ratio (SNR hereafter) to these galaxy images. For each galaxy image, we generate an image of pure noise whose statistical property is the same as the background noise on the galaxy image.

\begin{figure}[!t]
\centering
\includegraphics[width=6cm,height=8cm]{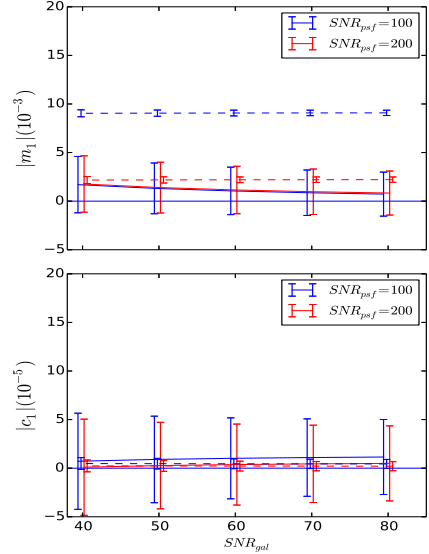}
\caption{The horizontal axis refers to the galaxy SNR, and the vertical axis are the multiplicative bias $m_{1}$ (upper panel) and the additive bias $c_{1}$ (lower panel) defined in eq.(\ref{shear_error}). '$SNR_{psf}$' refers to SNR for PSF. The solid and dashed curves are the results from our new method and ZLF15 respectively. }\label{fig_noise_psf_outcome} 
\end{figure}

In our image processing pipeline we cut off the galaxy and PSF images with circular top-hat functions of radii equal to $7$ times the half-light radii of the sources. The shear recovery accuracy are quantified by the multiplicative and addictive biases defined in eq.(\ref{shear_error}).  In fig.[\ref{fig_noiserevision_outcome}], we plot the results with our new method and ZLF15 for galaxies of different SNR's. The figure shows that both our new method and ZLF15 works well in the presence of background noise and source Poisson noise, though the statistical uncertainty of our new method is generally larger than that of ZLF15. 

The new method allows the PSF to have noise. Fig.[\ref{fig_noise_psf_outcome}] demonstrates that our new method works well even when the PSF has noise. In comparison, noise on PSF can cause systematic bias in ZLF15. The setup of the experiment is the same as the previous one, except that we add uncorrelated background Gaussian noise to both galaxy and PSF, and no source Poisson noise is added. Note that in the new method, for every noisy PSF image, we also need to generate a companion image of background noise. 

\section{Summary and Discussion}
\label{S-Summary}

Auto-convolved galaxy images can be used for shear estimation because it transforms similarly as the intrinsic image under the distortion matrix of cosmic shear (see eq.(\ref{shear auto-convolved})). The PSF effect can be corrected by subtracting the quadrupole moments of the auto-convolved PSF directly. The effect of noise can be statistically removed using neighbouring images of only background, similar to what is done in ZLF15. The PSF image is also allowed to contain noise, the effect of which can be similarly removed using neighbouring background images, as shown in \S\ref{noise_c}. These convenient features of the new method are all due to the linearity of the relation between the shear estimators and the multipole moments of the auto-convolved galaxy/PSF images.

The major advantages of the new method are: 1) It does not make assumptions on the morphologies of the galaxy or PSF; 2) It has an accurate treatment of noise for both galaxy and PSF; 3) The centroid of the auto-convolved images are trivially identified; 4) It only requires the quadrupole moments of the galaxy/PSF images; 5) The image processing of the method is very fast, as it only involves Fast Fourier Transformation. 

We choose to use circular aperture to define the boundary of the galaxy image, and to determine the proper aperture radii for galaxies of different Sersic profiles. The results show that the required aperture size (in terms of galaxy's half-light radius) for the new method is more strongly dependent on the galaxy morphology than the method of ZLF15. We also show that the new method works well in the presence of noise, albeit with a larger statistical error than ZLF15. These issues remain to be improved. A possible solution to these problems is to downweight the contribution of pixels at large distances from the image center. However, it seems that a nontrivial weighting function (comparing to our current circular top-hat function) neccessarily requires more complicated PSF correction procedures, as discussed in many other works. This is a possible direction for the future development of this method.

\acknowledgments{The authors thank Eiichiro Komatsu, Liping Fu, Guoliang Li, Zuhui Fan, Pengjie Zhang for useful discussions. JZ is supported by the national science foundation of China (Grant No. 11273018, 11433001), the national basic research program of China (Grant No. 2013CB834900, 2015CB857001), the national “Thousand Talents Program” for distinguished young scholars, a grant(No.11DZ2260700) from the Office of Science and Technology in Shanghai Municipal Government.}

\appendix

\section{A. Estimator with Quadrupole moments}
\label{App-PR}
Firstly, we substitute eq.(\ref{shear equation}) into eq.(\ref{PSF convolution})
\begin{equation}
f_o(\vec{x_o})=|\textbf{A}|\int{w(\vec{x_o}-\textbf{A}\vec{x_s})f_s(\vec{x_s})d^2 x_s}.
\end{equation}
Without loss of generality, we set $\kappa=0$, and perform the following calculation to the first order accuracy in shear. By defining $\vec{x_o}=\textbf{A}\vec{x}$, we can expand $w[\textbf{A}(\vec{x}-\vec{x_s})]$ and get
\begin{equation}\label{app-3}
\begin{split}
&f_o(\vec{x_o})=\int wf(\vec{x_s})d^2x_s\\
&+g_1[\int (x-x_s)\frac{\partial{}w}{\partial{}x}f_s(\vec{x_s})d^2x_s-\int(y-y_s)\frac{\partial{}w}{\partial{}y}f_s(\vec{x_s})d^2x_s]\\
&+g_2[\int(y-y_s)\frac{\partial{}w}{\partial{}x}f_s(\vec{x_s})d^2x_s+\int (x-x_s)\frac{\partial{}w}{\partial{}y}f_s(\vec{x_s})d^2x_s]
\end{split}
\end{equation}
where $w$ is a simple notation of $w(\vec{x}-\vec{x_s})$.\\
The moments of intrinsic galaxy image is defined as:($i,j=0,1,2$)
\begin{equation}
Q_s^{ij}=\int{x^iy^jf_s(\vec{x})d^2x}.
\end{equation}

From eq.(\ref{app-3}), we can get
\begin{equation}\label{app--4}
\begin{split}
\int x^2 f_o(\vec{x_o})d^2x&=R^{20}Q_s^{00}+R^{00}Q_s^{20}-2g_1R^{20}Q_s^{00}-2g_2R^{11}Q_s^{00}\\
\int y^2 f_o(\vec{x_o})d^2x&=R^{02}Q_s^{00}+R^{00}Q_s^{02}+2g_1R^{02}Q_s^{00}-2g_2R^{11}Q_s^{00}\\
\int xy f_o(\vec{x_o})d^2x&=R^{11}Q_s^{00}+R^{00}Q_s^{11}-g_2R^{20}Q_s^{00}-g_2R^{02}Q_s^{00}\\
\int f_o(\vec{x_o})d^2x&=R^{00}Q_s^{00}
\end{split}
\end{equation}
We can further write the moments of observed galaxy using the redefined coordinates $(\vec{x})$
\begin{equation}\label{app-8}
\begin{split}
Q_o^{20}&=\int[(1+2g_1){x}^2+2g_2xy]f_o(\vec{x_o})d^2x\\
Q_o^{02}&=\int[(1-2g_1){y}^2+2g_2xy]f_o(\vec{x_o})d^2x\\
Q_o^{11}&=\int[g_2({x}^2+{y}^2)+xy]f_o(\vec{x_o})d^2x\\
\end{split}
\end{equation}
Substitute eq.(\ref{app--4}) into eq.(\ref{app-8}), we get
\begin{equation}
\begin{split}\label{app--2}
(Q_o^{20}-Q_o^{02})R^{00}-(R^{20}-R^{02})Q_o^{00}&=(R^{00})^2(Q_s^{20}-Q_s^{02})+2g_1(R^{00})^2(Q_s^{20}+Q_s^{02})\\
Q_o^{11}R^{00}-R^{11}Q_o^{00}&=(R^{00})^2Q_s^{11}+g_2(R^{00})^2(Q_s^{20}+Q_s^{02})\\
(Q_o^{20}+Q_o^{02})R^{00}-(R^{20}+R^{02})Q_o^{00}&=(R^{00})^2(Q_s^{20}+Q_s^{02})+2g_1(R^{00})^2(Q_s^{20}-Q_s^{02})+4g_2(R^{00})^2Q_s^{11}
\end{split}
\end{equation}

Statistically, the intrinsic galaxies are isotropic so $<Q_s^{20}-Q_s^{02}>$ and $<Q_s^{11}>$ vanishes. Average both side of eq.(\ref{app--2}), it is straightforward to derive eq.(\ref{Estimator Revised}).

\section{B. Estimator with Autoconvolution}
\label{App-AC}
If the centroid of galaxy is $\vec{d}$ ($\vec{d}=(d_1,d_2)$) offsets from the origin, the moments measured from the centroid of observed galaxy should change to
\begin{equation}\label{qm QdR}
\begin{split}
Q_o^{ij}&=\int{(x_o-d_1)^i(y_o-d_2)^jf_o(\vec{x_o})d^2x_o}\\
d_1&=\frac{Q_d^{10}}{Q_d^{00}},~~d_2=\frac{Q_d^{01}}{Q_d^{00}}
\end{split}
\end{equation}
where $Q_d^{ij}=\int{x_o^iy_o^jf_o(\vec{x}_o)d^2x_o} $ are the moments measured from the origin of the coordinates (which is not in accordance with the centroid).
According to eq.(\ref{qm QdR}), we can write $Q_o^{ij}$ as function of $Q_d^{ij}$
\begin{equation}
\begin{split}
Q_o^{00}&=Q_d^{00}\\
Q_o^{20}&=Q_d^{20}-\frac{(Q_d^{10})^2}{Q_d^{00}}\\
Q_o^{02}&=Q_d^{02}-\frac{(Q_d^{01})^2}{Q_d^{00}}\\
Q_o^{11}&=Q_d^{11}-\frac{Q_d^{10}Q_d^{01}}{Q_d^{00}}
\end{split}
\end{equation}
The relationship between $Q_a^{ij}$ and $Q_d^{ij}$ are:
\begin{equation}
\begin{split}
Q_a^{00}&=Q_d^{00}Q_d^{00}\\
Q_a^{20}&=2[Q_d^{00}Q_d^{20}-(Q_d^{10})^2]\\
Q_a^{02}&=2[Q_d^{00}Q_d^{02}-(Q_d^{01})^2]\\
Q_a^{11}&=2[Q_d^{00}Q_d^{11}-Q_d^{10}Q_d^{01}]
\end{split}
\end{equation}
So the moments of autoconvolved galaxy are
\begin{equation} \label{Ap2 --1}
\begin{split}
Q_a^{00}=Q_o^{00}Q_o^{00}\\
Q_a^{20}=2Q_o^{00}Q_o^{20}\\
Q_a^{02}=2Q_o^{00}Q_o^{02}\\
Q_a^{11}=2Q_o^{00}Q_o^{11}
\end{split}
\end{equation}
We can get similar equation for PSF: $R_a^{00}=R^{00}R^{00}$, $R_a^{20}=2R^{00}R^{20}$, $R_a^{02}=2R^{00}R^{02}$, $R_a^{11}=2R^{00}R^{11}$.
Substitute them into eq.(\ref{app--2}) and we could get
\begin{equation}
\begin{split}\label{b--1}
(Q_a^{20}-Q_a^{02})R_a^{00}-(R_a^{20}-R_a^{02})Q_a^{00}&=2Q_o^{00}(R^{00})^3(Q_s^{20}-Q_s^{02})+4g_1Q_o^{00}(R^{00})^3(Q_s^{20}+Q_s^{02})\\
Q_a^{11}R_a^{00}-R_a^{11}Q_a^{00}&=2Q_o^{00}(R^{00})^3Q_s^{11}+2g_2Q_o^{00}(R^{00})^3(Q_s^{20}+Q_s^{02})\\
(Q_a^{20}+Q_a^{02})R_a^{00}-(R_a^{20}+R_a^{02})Q_a^{00}&=2Q_o^{00}(R^{00})^3(Q_s^{20}+Q_s^{02})+4g_1Q_o^{00}(R^{00})^3(Q_s^{20}-Q_s^{02})+8g_2Q_o^{00}(R^{00})^3Q_s^{11}
\end{split}
\end{equation}
Average both side of eq.(\ref{b--1}), we can get eq.(\ref{Estimator Centroid}).

\end{document}